\documentclass[usenatbib]{mnras}
\usepackage{amsmath}
\usepackage{graphicx}
\usepackage{amssymb}
\usepackage{url}

\usepackage{natbib}

\begin{document}

\title[Hierarchical BH triples in young star clusters]{Hierarchical black hole triples in young star clusters: impact of Kozai-Lidov resonance on mergers}
\author[Kimpson et al.]{Thomas O. Kimpson$^{1}$\thanks{E-mail: tomkimpson@gmail.com}, 
 Mario Spera$^{2}$, Michela Mapelli$^{2,3}$, Brunetto M. Ziosi$^{2}$ \\
$^{1}$Durham University, Department of Physics, South Road, Durham, DH1 3LE, UK\\
$^{2}$INAF, Osservatorio Astronomico di Padova, Vicolo dell'Osservatorio 5, I-35122, Padova, Italy\\
$^{3}$INFN, Milano Bicocca, Piazza della Scienza 3, I--20126, Milano, Italy}

\maketitle \vspace {7cm}
\bibliographystyle{mnras}

\begin{abstract}
Mergers of compact object binaries are one of the most powerful sources of gravitational waves (GWs) in the frequency range of second-generation ground-based gravitational wave detectors (Advanced LIGO and Virgo). Dynamical simulations of young dense star clusters (SCs) indicate that $\sim{}27$ per cent of all double compact object binaries are members of hierarchical triple systems (HTs). In this paper, we consider  570 HTs composed of three compact objects (black holes or neutron stars) that formed dynamically in $N$-body simulations of young dense SCs. We simulate them for a Hubble time with a new code based on the Mikkola's algorithmic regularization scheme, including the 2.5 post-Newtonian term. We find that  $\sim{}88$ per cent of the simulated systems develop Kozai-Lidov (KL) oscillations. KL resonance triggers the merger of the inner binary in three systems (corresponding to $0.5$ per cent of the simulated HTs), by increasing the eccentricity of the inner binary. Accounting for KL oscillations leads to an increase of the total expected merger rate  by $\approx{}50$ per cent. All binaries that merge because of KL oscillations were formed by dynamical exchanges (i.e. none is a primordial binary) and have chirp mass $>20$ M$_\odot$.  This result might be crucial to interpret the formation channel of the first recently detected GW  events.
\end{abstract}


\section{Introduction}
A large fraction of stars ($\sim{}10$ per cent, \citealt{raghavan2010,riddle2015}) are found to be a part of hierarchical triple systems (HTs) whereby the inner binary is orbited by a tertiary body with a much larger semi-major axis. HTs exhibit interesting dynamical effects; if the tertiary body is inclined with respect to the inner binary it can induce oscillations in the orbital parameters of the inner binary, the so-called eccentric Kozai-Lidov (KL) resonance \citep{kozai1962,lidov1962}. Notably, the tertiary body causes the eccentricity of the inner binary to oscillate between two extremes, with a maximum value given by $e_{\rm max} \approx \sqrt{1-5/3 \, \cos{}^2 i}$, where $i$ is the inclination of the tertiary orbit with respect to the inner binary (in the limit of a test particle secondary when the three-body Hamiltonian is expanded to quadrupole order). These oscillations occur over the timescale \citep{holman1997,kiseleva1998,antognini2014} 
\begin{equation} \label{eq:KL}
  T_{\rm K} = \frac{2P_t^2}{3\,{}\pi \,{}P_b} (1-e_t^2)^{3/2}\,{} \frac{m_b+m_t}{m_t}, 
\end{equation} 
where $P$ is the period, $e$ the eccentricity, $m$ the mass and the subscripts $b$ and $t$ denote the inner binary pair and tertiary, respectively. 

If the inner binary system is composed of two compact objects, KL cycles can drive the inner binary to merge via gravitational wave (GW) emission. The timescale for such a merger can be approximated as \citep{peters1964}
\begin{eqnarray}\label{eq:peters}
 t_{\rm GW} \approx 2 \, \, {\rm Gyr} \left (\frac{a_b}{0.01AU} \right)^4 (1-e_b^2)^{7/2} \left( \frac{1\,{}M_{\odot}}{m_1} \right) \nonumber\\ \left( \frac{1\,{}M_{\odot}}{m_2} \right)\left( \frac{2\,{}M_{\odot}}{m_1+m_2} \right)
\end{eqnarray}
 where $a_b$ is the semi-major axis of the inner binary, $e_b$ the eccentricity of the binary, and $m_1$ and $m_2$ the respective masses of the two members of the binary pair. It is therefore clear that should resonance effects lead to the eccentricity of the inner binary to approach unity, the binary may merge in a very short time. %It is therefore clear that resonance effects may lead the eccentricity of the inner binary to approach unity, and the binary can merge in a very short time.

KL cycles can trigger mergers of compact-object binaries (e.g. \citealt{miller2002,naoz2013,antonini2012,antonini2014,antognini2014,antognini2015,antognini2016,antonini2016}), enhance the formation of blue straggler stars \citep{naoz2014,antonini2016}, affect the orbits of planets (e.g. \citealt{martin2016}), and  drive the formation of low-mass X-ray binaries \citep{naoz2015}. Moreover, KL cycles can enhance double white-dwarf mergers enough to explain the type~Ia SN rate \citep{thompson2011}.  These effects are particularly important in globular clusters \citep{antonini2016,naoz2015}, where the high central stellar density favours the formation of stellar triples. 

Quantifying the impact of KL resonance on the merger rate of compact objects is of extreme importance in light of the recent first direct detection of GWs by the Advanced LIGO detectors \citep{abbott2016,LIGO2016,LIGO2016b,GW151226a,GW151226b}. Two GW events have been observed by Advanced LIGO so far, GW150914 \citep{abbott2016} and GW151226 \citep{GW151226a}. Both events are associated with the merger of double black hole (BH) systems. A third GW signal, LVT151012, might be associated with the merger of another double BH system, but has low significance.

In this paper, we focus on young dense star clusters (SCs) and open clusters, and we investigate the impact of KL resonances on the merger of BH binaries. We re-simulate HTs that have been found to form self consistently in $N$-body simulations \citep{ziosi2014}, accounting for the effects of KL oscillations so as to discern the effect of KL resonance on the merger rate of BH-BH systems. The paper is organized as follows: in Section 2 we outline the methods and simulations; in Section 3 we describe and discuss our results. %The implications for gravitational wave emission are discussed in Section 4 and 
 The conclusions are presented in Section~4.

\section{Numerical methods and setup}
The initial conditions were drawn from a set of 600 direct $N$-body realisations of young SCs, already discussed in \cite{mapelli2014} and in \cite{ziosi2014}. We extracted 570 HTs composed of three BHs or two BHs and a neutron star (NS) from these simulations and then re-simulated them with a regularized algorithm  including the 2.5 post-Newtonian (PN) term, but not the 1PN term. In the following, we discuss the details of the initial conditions, of the regularized code, and of the simulations.

\subsection{The initial conditions}
The 600 direct $N$-body simulations we start from were run using the {\sc starlab}\footnote{\tt http://www.sns.ias.edu/$\sim{}$starlab/} public software environment \citep{portegieszwart2001}. {\sc kira}, the direct $N$-body integrator included in {\sc starlab}, implements a Hermite 4th order integration algorithm \citep{makino1992} and a neighbours--perturbers scheme  to integrate tight binaries and multiple  systems.  This means that each binary (or multiple system) is treated in a different way according to the strength of the interactions with other stars. If the binary is unperturbed (i.e. it has no strong perturbers according to the criterion discussed in  \citealt{portegieszwart2001}), it is integrated analytically as two-body motion: it is seen by the other stars in the system as a point mass. If the binary has close perturbers, it is resolved into its components and integrated through direct summation. No regularization schemes are applied.

We included stellar and binary evolution, using the modified version of {\sc starlab} described in \cite{mapellibressan2013} and \cite{mapelli2016}. Stars evolve in radius, temperature and luminosity at different metallicities according to the polynomial fitting formulae by \cite{hurley2000}. A treatment of stellar winds is included for both  main sequence  and post-main sequence stars (\citealt{portegieszwart1996,vink2001,vink2005,belczynski2010}; see \citealt{mapelli2013} for a complete description).

The formation of stellar remnants is implemented as described in \cite{mapelli2013}. In particular, BH masses at various metallicities follow the distribution described in fig. 1 of \cite{mapelli2013} (see also \citealt{fryer1999,fryer2001, fryer2012,mapelli2009,mapelli2010,spera2015,spera2016}). If the final mass $m_{\rm fin}$ of the progenitor star (i.e. the mass bound to a star immediately before the collapse) is $>40$ M$_\odot$, we assume that the supernova (SN) fails and that the star collapses quietly to a BH. 
 The mass of a BH born from direct collapse is similar to the final mass of the progenitor star. Thus, BHs with mass  up to $\sim{}80$ M$_\odot$ ($\sim{}40$ M$_\odot$) can form if the metallicity of the progenitor is $Z\sim{}0.01$ Z$_\odot$ ($Z\sim{}0.1$ Z$_\odot$). 

 NSs are assumed to receive a natal kick drawn from the distribution of \cite{hartman1997}. BHs that form from quiet collapse are assumed to receive no natal kick \citep{fryer2012}. For BHs that form from a SN explosion, the natal kicks were drawn from the same distribution as NSs but scaled as $m_{\rm NS}/m_{\rm BH}$ (where $m_{\rm NS}=1.3$ M$_\odot{}$ is the typical mass of a NS, while $m_{\rm BH}$ is the mass of the considered BH), to preserve linear momentum.

The 600 runs we employ to derive the initial conditions include $N=5500$ particles each, corresponding to an average total mass $M\sim{}3.5\times{}10^3$ M$_\odot{}$. The SCs are modelled as King models \citep{king1966} with virial radius $r_{\rm v}=1$ pc, core radius $r_{\rm c}\sim{}0.4$ pc, half-mass radius $r_{\rm hm}\sim{}0.8$ pc, and central dimensionless potential $W_0=5$. These values are typical of intermediate-mass SCs in the Milky Way, such as the Orion Nebula Cluster \citep{portegieszwart2010}. We generate star masses according to a Kroupa initial mass function (IMF, \citealt{kroupa2001}) with minimum mass $m_{\rm low}=0.1$ M$_\odot$ and maximum mass $m_{\rm up}=150$ M$_\odot$. We include a fraction $f_{\rm PB}=0.18$ of stars in primordial binaries. This is a lower limit to the binary fraction observed in young SCs (e.g. \citealt{li2013}), but higher primordial binary fractions are prohibitive for the integration time. Additional binaries form by dynamical encounters. We consider three different metallicities: $Z=$ 1, 0.1, and 0.01 Z$_\odot$, respectively (here we assume Z$_\odot{}=0.02$). We simulate 200 different realizations for each of these metallicities, to increase the statistics and filter out statistical fluctuations. Each SC has been simulated in isolation for $t=100$ Myr. 

\cite{ziosi2014} analysed the formation of BH-BH binaries in these 600 simulations. They found that the vast majority ($\sim{}97$ per cent) of BH-BH binaries come from dynamical exchanges. Thus, SC dynamics is extremely important for the formation of BH-BH binaries (see also  \citealt{downing2010,downing2011,oleary2006,sadowski2008,clausen2013,morscher2015,rodriguez2015,oleary2016,chatterjee2016,giersz2015,mapelli2016,hurley2016} for similar conclusions). However, most BH-BH binaries are too loose to emit detectable gravitational waves: only 7 out of 2096 BH-BH binaries are expected to merge within a Hubble time in the sample of \cite{ziosi2014}. This implies an expected merger rate of a few $\times{}10^{-3}$ Gpc$^{-3}$ yr$^{-1}$.  The minimum merger rate consistent with the two detected events, GW150914  and GW151226, is $\sim{}9\times{}10^{-3}$ Gpc$^{-3}$ yr$^{-1}$ \citep{abbott2016,GW151226a}.

570 HTs composed of three BHs or two BHs and a NS form in the simulations analysed by \cite{ziosi2014}, approximately one per simulated SC. Triples hosting NSs are a minority: the inner binary is a BH-NS binary only in 5 cases, while the NS is the tertiary body in 3 additional systems.  In contrast, systems composed of three BHs are a striking majority ($\sim{}99$ per cent of all simulated compact-object HTs). Since 2096 BH-BH binaries form in all simulations by \cite{ziosi2014}, this means that $\sim{}27$ per cent of all BH-BH binaries went through the formation of a HT with a third BH (or NS) at least once during the simulations. This indicates that the formation of triple BH systems is a common feature in SCs.

%%%%%%%%%%%%%%%%%%%%%%%%%%%%%%%%FIG 1%%%%%%%%%%%%%%%%%%%%%%%%%%%%%%%%%%%%%%%
\begin{figure*}
\begin{center}
	\includegraphics[width=15cm]{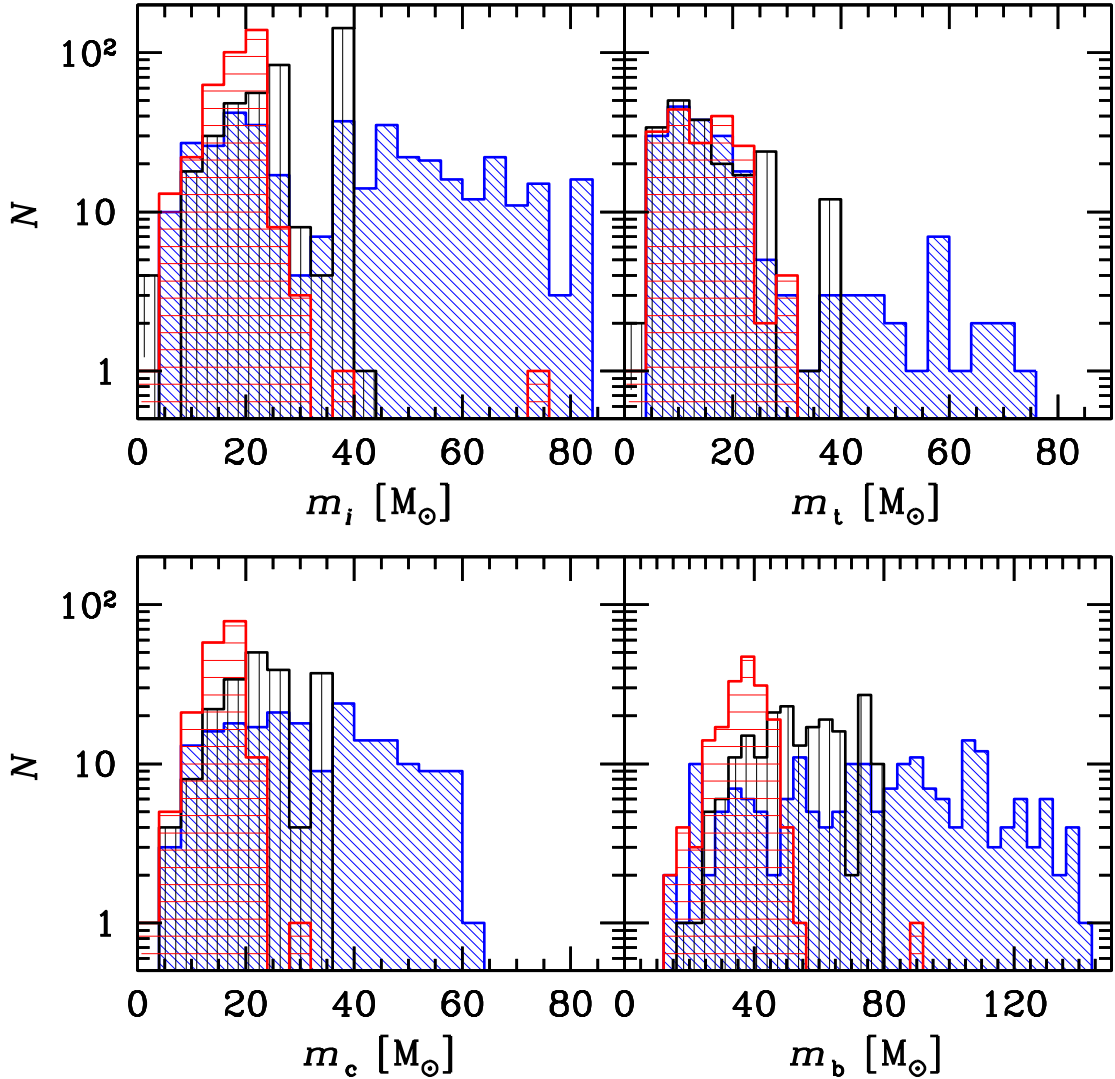} %WAS masses.eps
	\caption{\label{fig:fig1}
In the top row, the left-hand panel shows the distribution of the masses of the members of the inner binary ($m_i$, with $i=1,\,{}2$), while the right-hand panel shows the distribution of the masses of the tertiary ($m_{\rm t}$). In the bottom row, the left-hand panel and the right hand panel show the chirp mass $m_{\rm c}=(m_1\,{}m_2)^{3/5}\,{}(m_1+m_2)^{-1/5}$ and the total mass $m_{\rm b}=m_1+m_2$ of the inner binary, respectively. In all panels, diagonally hatched blue histogram: $Z=0.01$ Z$_\odot$; vertically hatched black histogram: $Z=0.1$ Z$_\odot$; horizontally hatched red histogram: $Z={\rm Z}_\odot$.
}
\end{center}
\end{figure*}
%%%%%%%%%%%%%%%%%%%%%%%%%%%%%%%%%%%%%%%%%%%%%%%%%%%%%%%%%%%%%%%%%%%%%%%%%%%%

The dynamical integrator included in {\sc starlab} is not sufficiently accurate to evolve HTs accounting for KL oscillations,  because no regularization scheme is implemented in {\sc kira}. This issue is particularly severe for high eccentricity binaries, such as those where KL oscillations are effective. Moreover, PN terms might be very important in the final evolution of some KL systems. Thus, we selected all triple BH systems from our {\sc starlab} simulations, and we re-simulated them with a new regularized $N$-body code including the 2.5PN term.

\subsection{The regularization code and post-Newtonian terms}
All HTs are evolved by means of a fully regularized $N$-body code that implements the Mikkola's algorithmic regularization (MAR, \citealt{mikkola1999a,mikkola1999b}). This code is particularly suitable for studying the dynamical evolution of few-body systems in which strong gravitational encounters are very frequent and the mass ratio between the interacting objects is large. The MAR scheme removes the singularity of the two-body gravitational potential for $r\rightarrow{}0$, by means of a transformation of the time coordinate (see \citealt{mikkola1999a} for the details). Our implementation uses a leapfrog scheme in combination with the Bulirsh-Stoer extrapolation algorithm \citep{stoer1980} to increase the accuracy of the numerical results. 

We also included the dissipative relativistic correction to the Newtonian forces that accounts for the quadrupole gravitational radiation. This correction, indicated as 2.5PN, comes from the post-Newtonian approximation theory
and  is of order $1/c^5$, where  $c$ is the speed of light (e.g. \citealt{kupi2006}). Since the 2.5PN term depends on the velocity of the particles, we modified the leapfrog scheme into a generalized midpoint method to maintain the correct symmetry of the integration scheme in the case of velocity dependent perturbations \citep{mikkola2006,mikkola2008}. In this way, we can successfully embed the integrator inside the Bulirsh-Stoer scheme to increase the accuracy of the integration.  Other post-Newtonian terms are not included in the current version of the code.

The code integrates the equations of motion employing relative coordinates by means of the so called chain structure. This change of coordinates reduces round-off errors significantly \citep{aarseth2003}.

At present, this code is a sub-module of the direct $N$-body code HiGPUs-R which
is still under development (Spera, in preparation; see \citealt{capuzzo2013}  for the current non-regularized version of HiGPUs). Still, it can be used as a stand-alone tool to study the dynamical evolution of few-body systems with very high precision.

\subsection{Simulations and {\it caveats}}\label{sec:simulations}
 We take each of the 570 HTs at the time it first appears in the $N$-body simulation. Then, we simulate it in isolation with the regularized code for a Hubble time ($t_{\rm H}=14$ Gyr).  To identify genuine KL oscillations in our simulations, we search for periodic peaks over a timescale of the same order as the theoretical Kozai timescale (equation~\ref{eq:KL}). If such peaks are found, KL oscillations are considered important.

Before analyzing the results of our simulations, we briefly discuss their main issues. Simulating the HTs in isolation, we neglect possible perturbers. Perturbers are expected to be important for binaries and multiple systems in dense SCs. A simple estimate of the three-body encounter rate (based on the geometrical cross-section of a binary corrected by gravitational focusing, e.g. \citealt{colpi2003}) gives
\begin{eqnarray}
R_{\rm en}=\frac{2\,{}\pi{}G\,{}m_{\rm b}\,{}a_{\rm b}\,{}n_\ast{}}{\sigma{}_\ast{}}\sim{}\nonumber{}\\0.1\,{}{\rm Myr^{-1}}\left(\frac{m_{\rm b}}{50\,{}{\rm M}_\odot}\right)\,{}\left(\frac{a_{\rm b}}{100\,{}{\rm AU}}\right)\,{}\left(\frac{n_\ast}{10^3 {\rm pc}^{-3}}\right)\,{}\left(\frac{5 {\rm km \,{}s}^{-1}}{\sigma{}_\ast}\right),
\end{eqnarray}
where $m_{\rm b}$ and $a_{\rm b}$ are the binary mass and semi-major axis; $n_{\ast}$ and $\sigma_\ast$ are the number density and velocity dispersion of stars close to the binary, respectively. If the SC survives tidal disruption for $\sim{}100$ Myr, a binary in its core will thus undergo $\sim{}10$ encounters. This value depends on the SC lifetime, on the local density and on the velocity dispersion in the neighborhoud of a binary system. 

Close encounters can either break the HT (whose stability is extremely precarious, see e.g. \citealt{aarseth2001}) or re-orient it. It is even possible that re-orientation favours the merger of the inner binary. Moreover, weak encounters
with distant perturbers have less dramatic effects but are far more frequent, and can also contribute to harden or soften binaries \citep{heggie1975}.

These {\it caveats} must be considered when interpreting the results of our work. A further study is needed, to investigate the impact of strong and weak perturbations. On the other hand, the HTs where the inner binary merges triggered by KL oscillations (see Section~\ref{sec:results}) are not affected by strong perturbers during the entire $N$-body simulation performed with {\sc starlab}.

Another important {\it caveat} is represented by relativistic effects. In our simulations, we include the 2.5PN term, but we neglect other post-Newtonian terms. However, general relativistic precession (described by 1PN) can stop eccentricity oscillations by destroying the Kozai resonance (e.g. \citealt{holman1997,hollywood1997,blaes2002,antonini2012}). KL oscillations are not suppressed by relativistic precession only if \citep{blaes2002} 
\begin{eqnarray}\label{eq:blaes}
\frac{a_{\rm t}}{a_{\rm b}}<910\,{}\left(\frac{a_{\rm b}}{1000\,{}{\rm AU}}\right)^{1/3}\,{}\left(\frac{m_{\rm b}}{50\,{}{\rm M}_\odot{}}\right)^{-1/3}\nonumber{}\\\times{}\,{}\left(\frac{2\,{}m_{\rm t}}{m_{\rm b}}\right)^{1/3}\, \left(\frac{1-e_{\rm b}^2}{1-e_{\rm t}^2}\right)^{1/2},
%\frac{a_{\rm t}}{a_{\rm b}}<34\,{}\left(\frac{a_{\rm b}}{10^{-2}{\rm pc}}\right)^{1/3}\,{}\left(\frac{m_{\rm b}}{2\times{}10^6{\rm M}_\odot{}}\right)^{-1/3}\nonumber\\\left(\frac{2\,{}m_{\rm t}}{m_{\rm b}}\right)^{1/3}\, \left(\frac{1-e_{\rm b}^2}{1-e_{\rm t}^2}\right)^{1/2},
\end{eqnarray}
where $a_{\rm b}$ ($a_{\rm t}$) is the semi-major axis of the inner (outer) binary, $e_{\rm b}$ ($e_{\rm t}$)   is the eccentricity of the inner (outer) binary, $m_{\rm b}$ is the total mass of the inner binary and $m_{\rm t}$ is the mass of the tertiary body. In the following analysis, we will use this simple analytic formalism to predict whether relativistic precession can influence the merger of simulated systems. %the system is then classified as Kozai.

%We then classify each system as either %(i) a system in which KL oscillations are sufficiently important to trigger the merger of the inner binary in a Hubble time; (ii) a system that undergoes KL oscillations, but the inner BH-BH binary does not merge within a Hubble time;  (iii) a system that undergoes a dynamical exchange (i.e. the tertiary body becomes member of the inner binary and, usually, the former member of the inner binary is ejected from the system); (iv) any other cases. These are all systems that do not undergo an exchange, and in which the oscillations do not match the Kozai timescale and/or their amplitude is $<1$ per cent of the eccentricity. %. systems where the inner binary splits). 

%%%%%%%%%%%%%%%%%%%%%%%%%%%%%%%%%%%%%%FIG 2%%%%%%%%%%%%%%%%%%%%%%%%%%%%%%%%%
\begin{figure*}
\begin{center}
	\includegraphics[width=15cm]{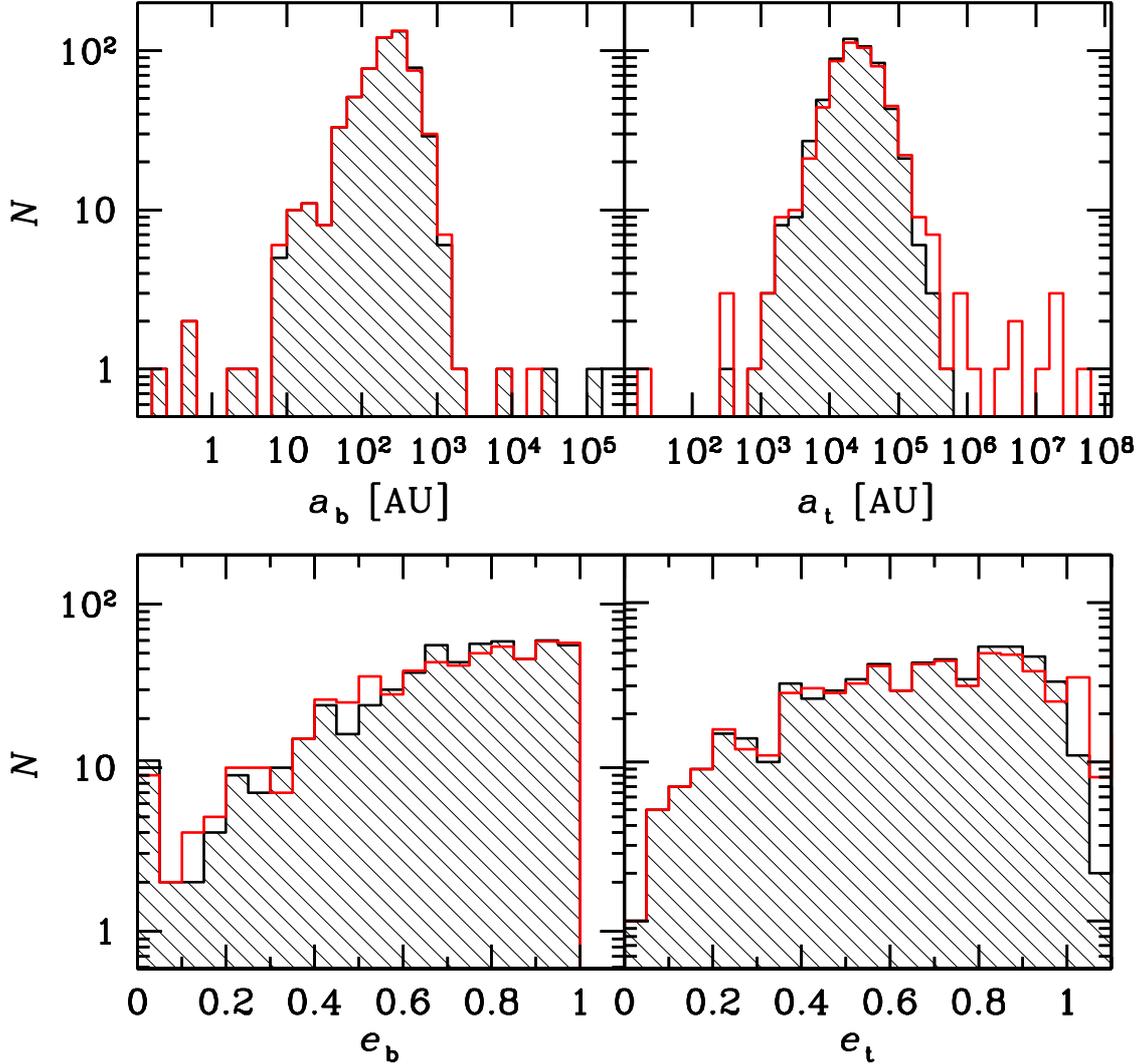} %was sma_ecc.eps}
	\caption{\label{fig:fig2}
In the top row, the left-hand and the right-hand panel show the distribution of the semi-major axis of the inner binary ($a_{\rm b}$) and of the outer binary ($a_{\rm t}$), respectively. In the bottom row, the left-hand and the right-hand panel show the distribution of the eccentricity of the inner binary ($e_{\rm b}$) and of the outer binary ($e_{\rm t}$), respectively. In all panels, the black diagonally hatched histogram refers to the initial conditions, while the  red open histogram indicates the end of the simulations.
}
\end{center}
\end{figure*}
%%%%%%%%%%%%%%%%%%%%%%%%%%%%%%%%%%%%%%%%%%%%%%%%%%%%%%%%%%%%%%%%%%%%%%%%%%%%%

\section{Results}\label{sec:results}
\subsection{Orbital properties of simulated HTs}
Figure~\ref{fig:fig1} shows the relevant masses of the simulated systems, divided by metallicity. The masses depend on the metallicity by construction, according to the adopted recipes of stellar evolution and stellar winds \citep{mapelli2013}. On average, the tertiary is slightly less massive than both members of the binary. In fact, if the tertiary was more massive than any member of the binary, a dynamical exchange would occur very fast, because the probability of an exchange strongly depends on the mass ratio of the involved bodies  (\citealt{mapelli2014} and references therein). A dynamical exchange leads to the ejection of one of the former members of the binary and prevents the formation of a HT.

The chirp mass of the inner binary, defined as $m_{\rm c}=(m_1\,{}m_2)^{3/5}(m_1+m_2)^{-1/5}$, where $m_1$ and $m_2$ are the masses of the two components of the binary, is particularly important for GWs, since it determines how fast the binary sweeps, or chirps, through a frequency band (the amplitude and the frequency of GWs scale as $m_{\rm c}^{5/3}$ and $m_{\rm c}^{-5/8}$, respectively). The range of chirp masses is $\sim{}5-25$ M$_\odot$, $\sim{}5-36$ M$_\odot$, and $\sim{}5-65$ M$_\odot$ at $Z=1$, 0.1 and 0.01 Z$_\odot$, respectively. Correspondingly, the range of total masses of the binary $m_{\rm b}=m_1+m_2$ is $\sim{}10-55$ M$_\odot$, $\sim{}10-80$ M$_\odot$, and $\sim{}10-150$ M$_\odot$ at $Z=1$, 0.1 and 0.01 Z$_\odot$, respectively. This range of chirp and total masses is consistent with the events that were recently detected by Advanced LIGO \citep{LIGO2016}.

Figure~\ref{fig:fig2} shows the distribution of initial eccentricity and semi-major axis of the inner and outer binary at the start and the end of the simulations, across all systems. We found no global change of the distribution of semi-major axis, consistent with the fact that KL resonance does not affect the energy of the binary. Only exchanges and PN terms can change the semi-major axis, but they affect only a small minority of our runs. The Kolmogorov-Smirnov (KS) test indicates that the probability that the initial and final semi-major axes are drawn from the same distribution is $\sim{}1$ and $\sim{}0.7$ for the inner and outer binary, respectively. The majority of inner binaries have a semi-major axis $a_{\rm b}\sim{}10^2-10^3$ AU, while the semi-major axis of the outer binary is much higher ($\sim{}10^4-10^5$ AU).
 
As to the eccentricity distribution, we find no apparent change of eccentricity distribution, consistent with the fact that KL oscillations do not change the average eccentricity in time. The KS test indicates that the probability that the initial and final eccentricities are drawn from the same distribution is $\sim{}0.4$ and $\sim{}0.2$ for the inner and outer binary, respectively. These probabilities are quite lower than for the semi-major axis, but we cannot reject the hypothesis that these distributions come from the same one.

\begin{figure}
\begin{center}
	\includegraphics[width=8cm]{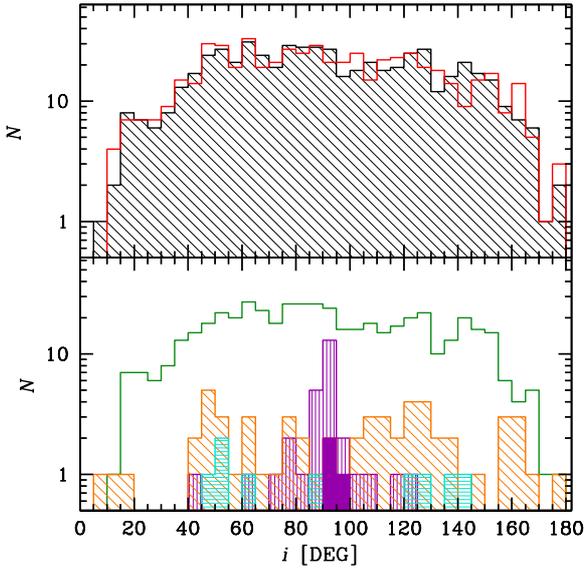}
	\caption{\label{fig:fig3}
 Distribution of the inclinations between the plane of the inner binary and that of the outer binary. Top panel: inclinations at the beginning (black diagonally hatched histogram) and at the end of the simulations (red open histogram) for all simulated systems. Bottom panel: initial inclinations of systems in class~A (dark green open histogram), class~B (turquoise horizontally hatched histogram), class~C (orange diagonally hatched histogram), class~A1 (violet vertically hatched histogram) and class~A2 (violet filled histogram). Class~A2 corresponds to the three merging systems. See Table~\ref{tab:table1} and Section~\ref{sec:statistics} for details on the classification of HTs in these groups.
}
\end{center}
\end{figure}

Finally, Figure~\ref{fig:fig3} shows the distribution of inclinations $i$ between the orbital plane of the inner and outer binary. The distribution is nearly flat and does not change significantly during the integration  (top panel of Fig.~\ref{fig:fig3}), consistently with the behaviour of KL resonance. The KS test indicates that the probability that the initial and final inclinations are drawn from the same distribution is $\sim{}1$. In Figures~\ref{fig:fig2} and \ref{fig:fig3}, we do not distinguish the three metallicities because, as already shown by \cite{ziosi2014}, the orbital properties do not depend on the metallicity significantly.

%%%%%%%%%%%%%%%%%%%%%%%%%%%%%%%%%%%%FIG 3%%%%%%%%%%%%%%%%%%%%%%%%%%%%%%%
%%%%%%%%%%%%%%%%%%%%%%%%%%%%%%%%%%%%%%%%%%%%%%%%%%%%%%%%%%%%%%%%%%%%%

\subsection{Statistics of KL systems}\label{sec:statistics}

%%%%%%%%%%%%%%%%%%%%%%%%%%%%%%%%%%%%FIG 4%%%%%%%%%%%%%%%%%%%%%%%%%%%%%%%%%%
\begin{figure}
\begin{center}
	\includegraphics[height=12cm]{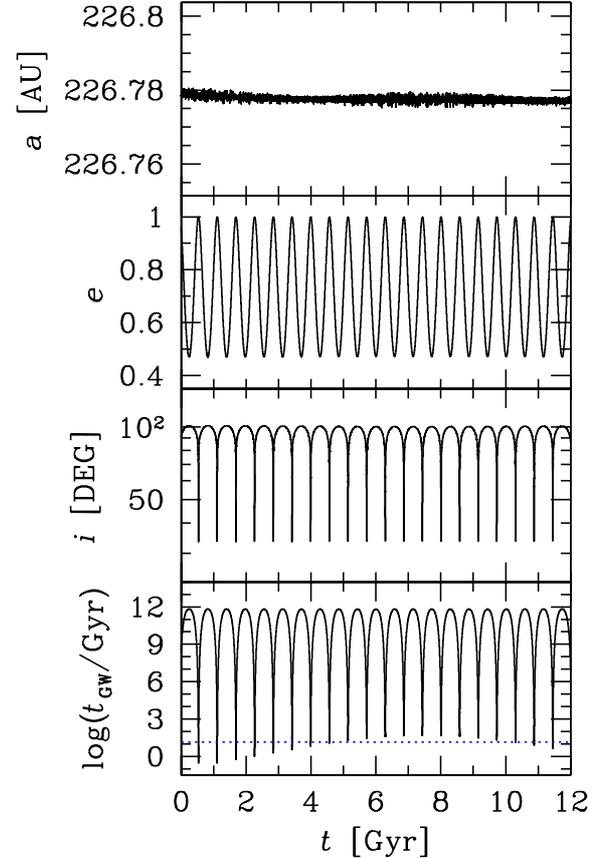}
	\caption{\label{fig:fig4}
 Properties of the inner binary of a HT showing regular KL cycles. From top to bottom: time dependence of semi-major axis  ($a_{\rm b}$), eccentricity  ($e_{\rm b}$), inclination with respect to the outer binary ($i$), and coalescence timescale $t_{\rm GW}$ (eq.~\ref{eq:peters}).  Blue dotted line in the bottom panel: Hubble time $t_{\rm H}$. This simulation was run with the 2.5PN term.}
\end{center}
\end{figure}
%%%%%%%%%%%%%%%%%%%%%%%%%%%%%%%%%%%%%%%%%%%%%%%%%%%%%%%%%%%%%%%%%%%%%%%%%%%

%%%%%%%%%%%%%%%%%%%%%%%%%%%%%%%%%%%%%TABLE 1%%%%%%%%%%%%%%%%%%%%%%%%%%%%%%%%%
\begin{table}
\begin{center}
\caption{\label{tab:table1}
Classification of the HTs.}
 \leavevmode
\begin{tabular}[!h]{c|ccccc}
\hline
			Class   & A & A1 & A2 & B & C \\
\hline 
			Frequency   & 502   &  30 & 3    & 10 &  58 \\ 
			Percentage  & 88.0 &   5.3 & 0.5 & 1.8 & 10.2 \\ 
\noalign{\vspace{0.1cm}}
\hline
\end{tabular}
\end{center}
\footnotesize{Class~A: systems where KL oscillations are effective; class~A1: sub-sample of class~A where $t_{\rm GW}<t_{\rm H}$ for a fraction of the KL cycle; class~A2: systems that merge during the simulations (sub-sample of class~A1); class~B: exchanges; class~C: other systems (where Kozai is not important).}
\end{table}
%%%%%%%%%%%%%%%%%%%%%%%%%%%%%%%%%%%%%%%%%%%%%%%%%%%%%%%%%%%%%%%%%%%%%%%%%%%%%%

 Table~\ref{tab:table1} summarizes our classification of the 570 simulated systems.  We classify each system as either (i) a system that undergoes KL oscillations (hereafter, class~A);  (ii) a system that undergoes a dynamical exchange, i.e. the tertiary body becomes member of the inner binary and, usually, the former member of the inner binary is ejected from the system (class~B); (iii) any other system (class~C). These are all systems that do not undergo an exchange, and in which the oscillations do not match the Kozai timescale and/or their amplitude is $<1$ per cent of the eccentricity. %. systems where the inner binary splits).

Among the systems that undergo KL oscillations (class~A) we then consider the sub-sample of systems for which $t_{\rm GW}<t_{\rm H}$ for at least a fraction of the KL cycle (where $t_{\rm GW}$ is defined in equation~\ref{eq:peters} and $t_{\rm H}$ is the Hubble time). These systems are labelled as class~A1. A further sub-sample of class~A1 is represented by systems that merge during the simulations as an effect of KL oscillations and GW emission (these will be indicated as class~A2).

From Table~\ref{tab:table1} it is apparent that a large fraction of HTs ($\sim{}88$ per cent) exhibit KL oscillations (class~A), whilst $\sim{}2$ per cent undergo an exchange (class~B). In $\sim{}10$ per cent of HTs KL resonance is not efficient (class~C).

Furthermore,  30 inner binaries, i.e $\sim{}5$ per cent of all simulated systems, have $t_{\rm GW}<t_{\rm H}$ during a fraction of their KL cycles, because the eccentricity becomes $\sim{}1$ (class~A1).  However, in 27 out of 30 systems (among which, the system shown in Figure~\ref{fig:fig4}), the time spent in the configuration with $t_{\rm GW}<t_{\rm H}$ is not sufficiently long to produce effective orbital decay by GW emission (which, in our simulations, is modelled as 2.5PN term). 

\begin{figure*}
\begin{center}
	\includegraphics[height=12cm]{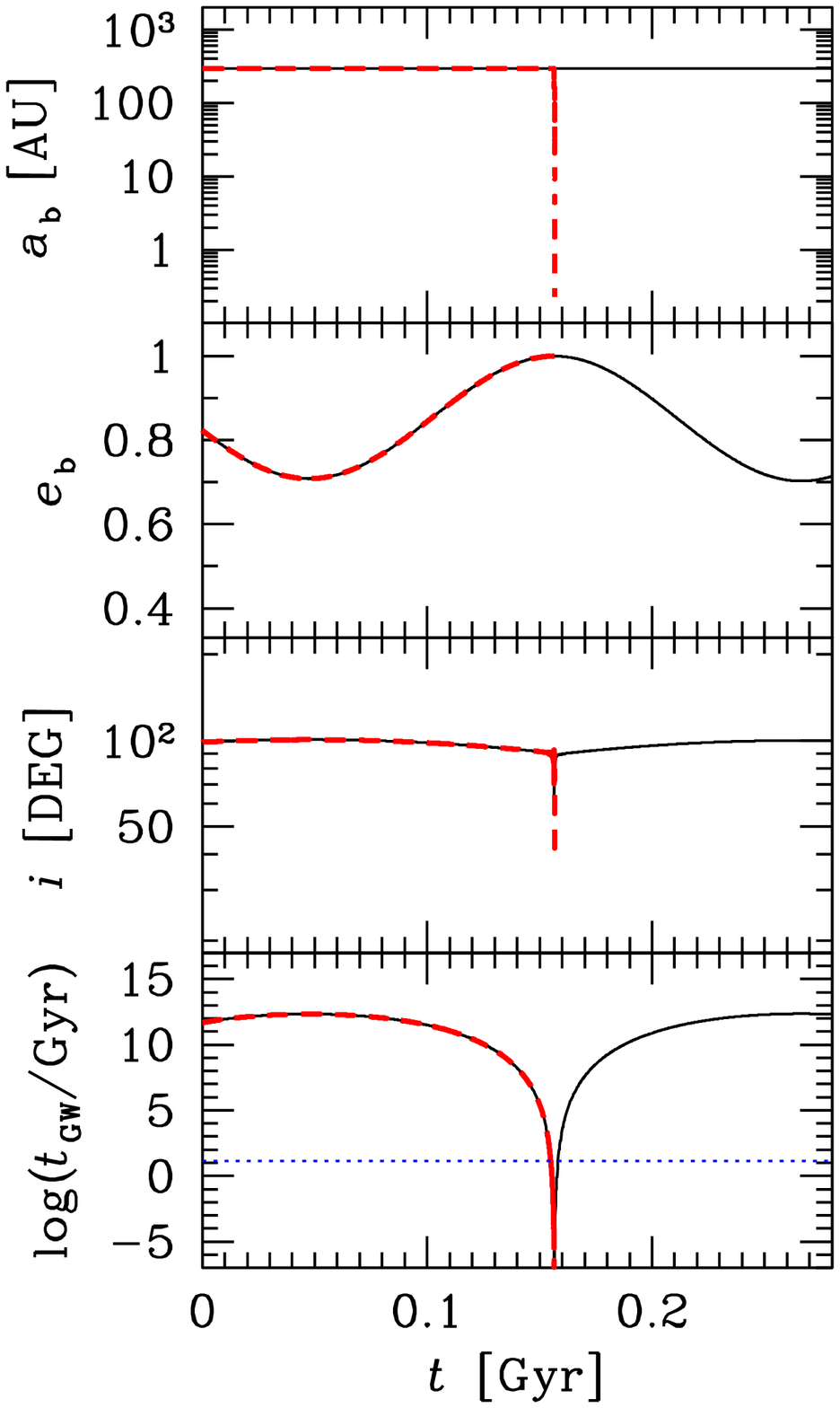}
	\includegraphics[height=12cm]{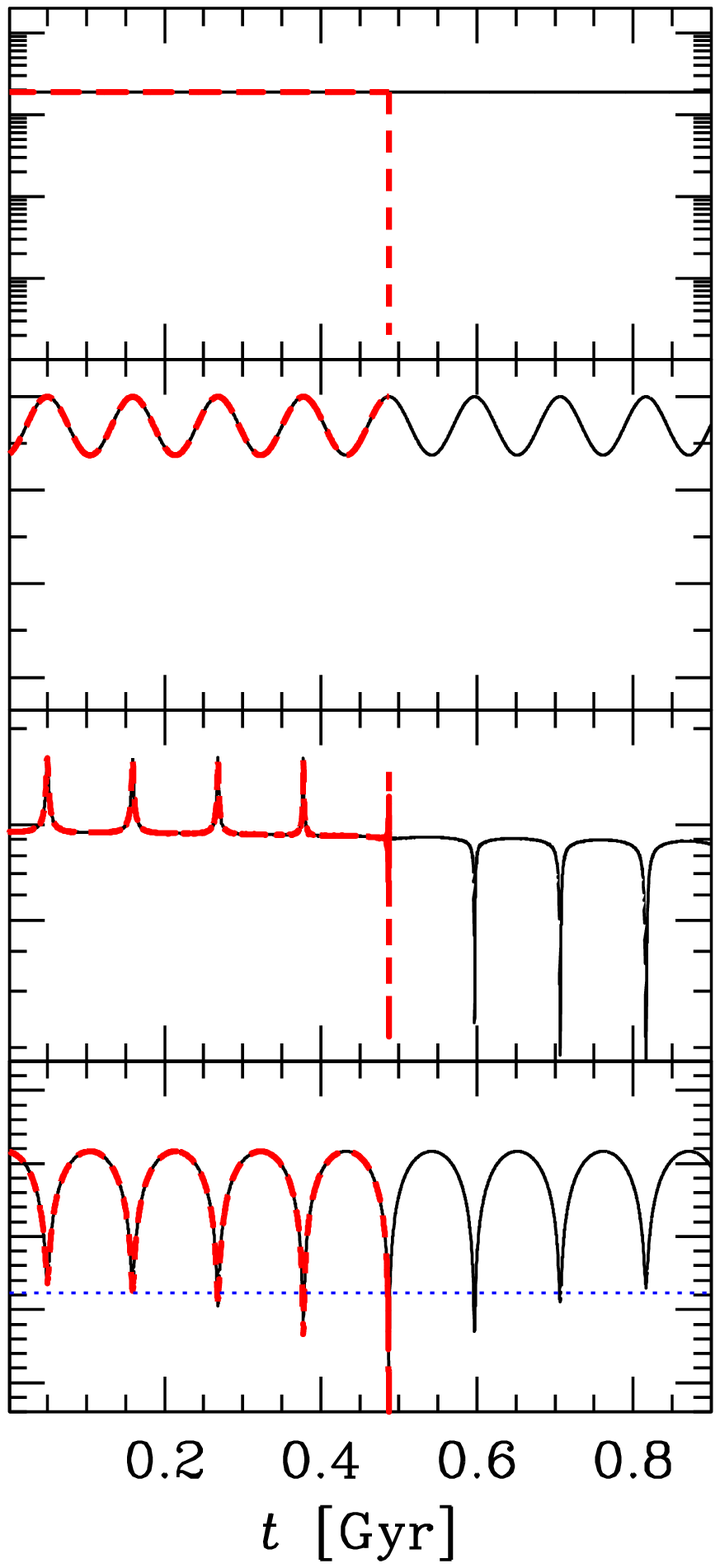}
	\includegraphics[height=12cm]{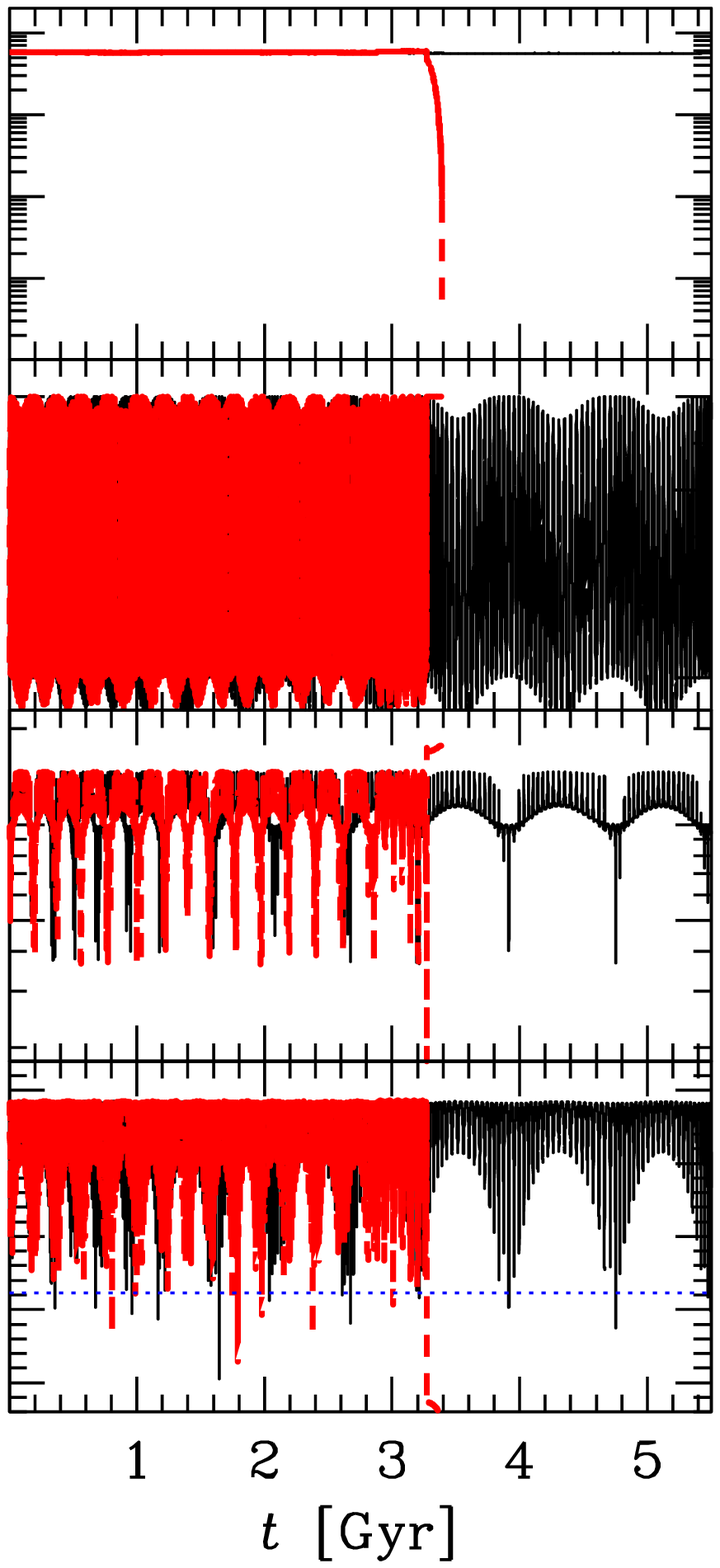}
	\caption{\label{fig:fig5}
 Properties of the inner binary in the three HTs merging within a Hubble time. For each system, from top to bottom: time dependence of semi-major axis ($a_{\rm b}$), eccentricity ($e_{\rm b}$), inclination ($i$) with respect to the outer binary, and coalescence timescale $t_{\rm GW}$  (eq.~\ref{eq:peters}). Red dashed line: simulation with 2.5PN term; black solid line: same simulation run without 2.5PN term. Blue dotted line in the bottom panels: Hubble time $t_{\rm H}$.
}
\end{center}
\end{figure*}

Only  $3$  inner binaries ($0.5$ per cent of the simulated systems) are observed to merge, as a consequence of KL oscillations (class~A2). All three merging systems are triple BH systems; no NS-BH system merges. 
The orbital decay induced by GW emission is effective in the three merging systems. Figure~\ref{fig:fig5} shows the time dependence of semi-major axis, eccentricity, inclination, and coalescence timescale ($t_{\rm GW}$) of the three merging systems. For all of them, we compare the evolution with the 2.5PN term to the evolution without 2.5PN term.

 The three merging systems (class~A2) all have  inclination $i\sim{}90$ DEG, as it is shown in the bottom panel of Figure~\ref{fig:fig3}. Similarly, most systems in class~A1 have inclination $i\sim{}90$ DEG. In contrast, exchanged (class~B) and `other' systems (class~C) tend to have either smaller or larger inclinations. This is consistent with the expectation that KL oscillations are maximally efficient for inclinations close to $\sim{}90$ DEG.

For the three merging systems we also checked the effect of relativistic precession by using equation~\ref{eq:blaes}. We found that the ratio $a_{\rm t}/a_{\rm b}$ is smaller than the critical value for relativistic precession to stop KL oscillations for the entire simulation, with the exception of the very last epoch of the merger phase (when $e_{\rm b}>0.99$ and when 2.5PN corrections are very effective, and $a_{\rm b}$ drops in few seconds). Thus, we expect that relativistic precession cannot stop the merger of these three systems.

%%%%%%%%%%%%%%%%%%%%%%%%%FIG 5%%%%%%%%%%%%%%%%%%%%%%%%%%%%%%%%%%%%%%%%%%%%%%%%%%
%%%%%%%%%%%%%%%%%%%%%%%%%%%%%%%%%%%%%%%%%%%%%%%%%%%%%%%%%%%%%%%%%%%%%%%%%%%%%

\subsection{Properties of merging systems}
Table~\ref{tab:table2} shows the properties of the three systems that merge in a Hubble time because of KL oscillations and GW emission (hereafter, we will call them `K16 merging systems', for brevity). For comparison, we list also the properties of the  events detected by Advanced LIGO \citep{abbott2016,GW151226a}, and those of the seven simulated BH-BH systems that were found to merge in a Hubble time by \cite{ziosi2014} (hereafter, `Z14 merging systems'). The seven Z14 merging systems come from the same set of simulations as the BH-BH systems we are considering in this paper, but their merger is not triggered by KL  resonance. %(we recall that the simulations of \citealt{ziosi2014} are not sufficiently accurate to integrate KL oscillations). 

In Table~\ref{tab:table2}, we also show the quantity $\tilde{t}_{\rm GW}$ (effective coalescence time), defined as
\begin{equation}\label{eq:tildetgw}
\tilde{t}_{\rm GW}=t_{\rm bin}+t_{\rm GW},
\end{equation}
where $t_{\rm bin}$ is the time elapsed from the formation of the binary till the end of the simulation, and $t_{\rm GW}$  the coalescence timescale defined in eq.~\ref{eq:peters}, estimated at the end of the simulation. For the Z14 merging systems $t_{\rm GW}>t_{\rm bin}$ (because \citealt{ziosi2014} simulated the evolution of SCs for 100 Myr).

 In the case of the three K16 merging systems, accounting for the 2.5PN term slows down the integration and makes the simulation stall: $t_{\rm bin}$ is the duration of the simulation till it stalls, and we evaluate $t_{\rm GW}$ at the stalling time. For the three K16 merging systems $t_{\rm GW}\lesssim{}1$ yr, hence $\tilde{t}_{\rm GW}\sim{}t_{\rm bin}$.

Even if the statistics of merging systems is low, we can do several interesting considerations. 
\begin{itemize}
\item[i)] In both \cite{ziosi2014} and this paper no BH-BH system  is expected to merge within a Hubble time from solar-metallicity progenitors. All merging systems form from stars with metallicity $\lesssim{}0.1$ Z$_\odot$. The  first  observed merger event, GW150914, is also thought to be associated with low metallicity ($<0.5$ Z$_\odot$), because of the large BH mass \citep{abbott2016}. 
 \item[ii)] The masses of K16 merging systems are generally higher than the masses of Z14 merging systems. In particular,  the mass of the primary BH ($39$ M$_\odot$) in the K16 merging system at $Z=0.1$ Z$_\odot$ is consistent with the mass of the primary BH in GW150914 ($36.2^{+5.2}_{-3.8}$ M$_\odot$, \citealt{abbott2016}). %, while the mass of the secondary ($18$ M$_\odot$) is  lower than that of the secondary BH in GW150914 ($29.1^{+3.7}_{-4.4}$ M$_\odot$). 
In contrast, the BH masses of most Z14 merging systems are fairly consistent with those of GW151226 (Table~\ref{tab:table2}).  %remarkably similar, although not consistent, with the ones of GW150914. 
\item[iii)] All three K16 merging  systems involve binaries that formed from a dynamical exchange, while  $\sim{}70$ per cent of Z14 merging systems occur in primordial binaries. Thus, dynamics is very important for all K16 merging systems. %Thus, mergers occurring in HTs are the consequence of dynamical interactions. 
\item[iv)] For all simulated merging systems, both Z14 and K16, $\tilde{t}_{\rm GW}<4$ Gyr, much shorter than the Hubble time.
\end{itemize}

\begin{figure}
\begin{center}
	\includegraphics[width=8cm]{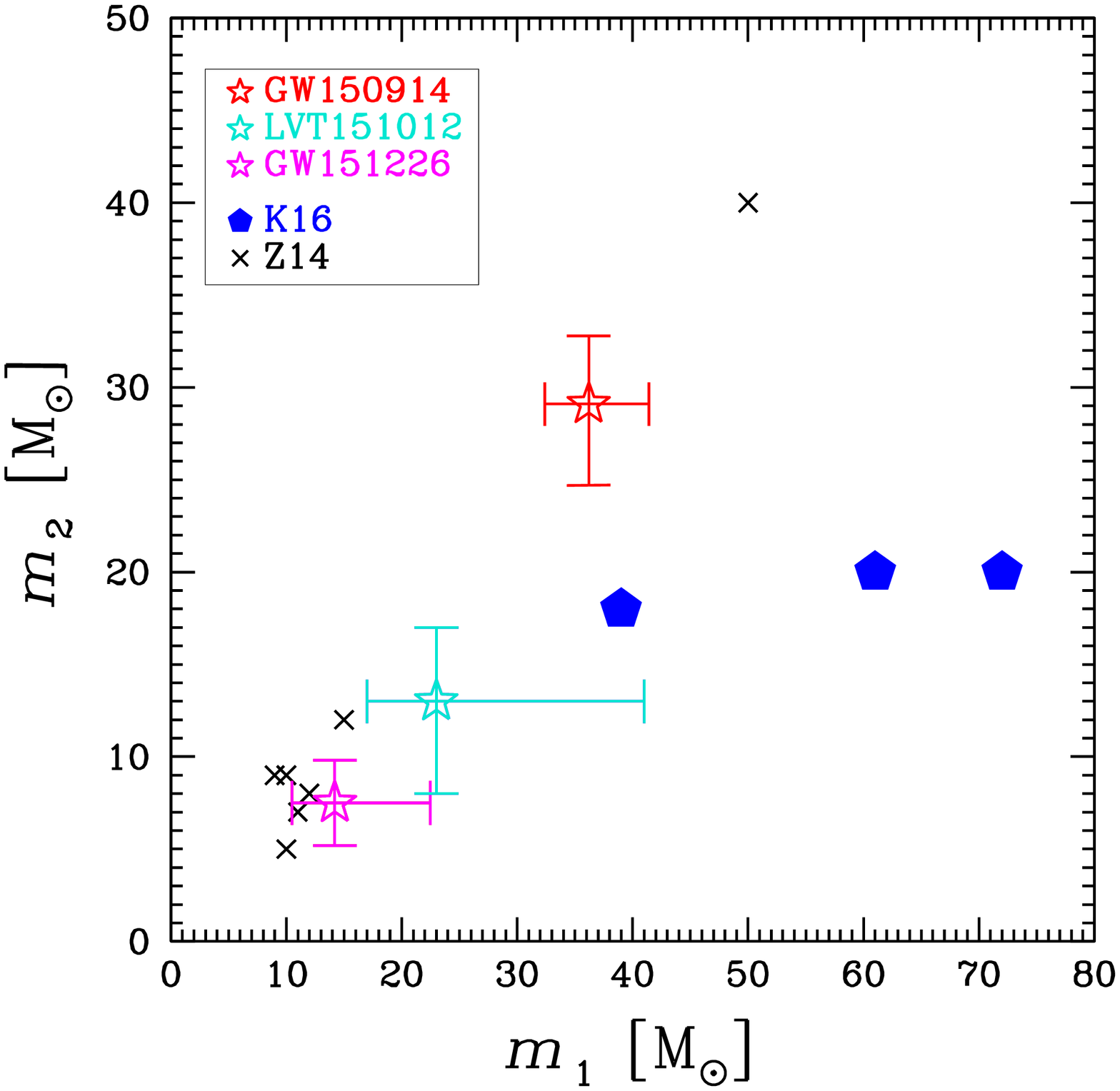}
	\caption{\label{fig:fig6}
Mass of the primary member of a merging binary ($m_1$) versus the mass of the secondary member ($m_2$). Black crosses: Z14 merging systems; blue filled pentagons: K16 merging systems; red  star with error bars: GW150914; magenta  star with error bars: GW151226; turquoise  star with error bars: LVT151012.}
\end{center}
\end{figure}

Figure~\ref{fig:fig6} shows the systems listed in Table~\ref{tab:table2} in the plane of $m_1,\,{}m_2$. All but one Z14 merging system have relatively low masses  (5--20 M$_\odot$),  matching the mass of GW151226. 

 The mass of one of the primary members in K16 matches the mass of  the primary member of GW150914, while the mass of the secondary members is slightly lower. The mass of the primary BH in K16 merging systems is always $\gtrsim{}40$ M$_\odot$. This might indicate that Kozai-induced mergers are skewed toward larger BH-BH masses. This consideration is consistent with the fact that HTs form dynamically, and dynamically born systems tend to have larger masses than primordial binaries (e.g. \citealt{mapelli2014} and references therein). However, our sample is far too small to draw any definitive conclusion. A large parameter-space investigation is requested to have statistically significant results.

%%%%%%%%%%%%%%%%%%%%%%%%%%%%%%%FIG 6%%%%%%%%%%%%%%%%%%%%%%%%%%%%%%%%%%%%%%
%%%%%%%%%%%%%%%%%%%%%%%%%%%%%%%%%%%%%%%%%%%%%%%%%%%%%%%%%%%%%%%%%%%%%%%%%%%%

\subsection{Impact on the merger rate}
From the results of Table~\ref{tab:table2} we can determine the merger rate for BH-BH binaries using a Drake-like equation \citep{esposito2015}:
\begin{eqnarray}\label{eq:drake} 
R\sim{} 5\,{}{\rm Gpc}^{-3}\,{}{\rm yr}^{-1}\,{}\left(\frac{t_{\rm life}}{10^8\,{}\text{yr}} \right)\,\left(\frac{\rho_{\rm SF}}{0.015\, \text{ M}_{\odot} \text{ yr}^{-1} \text{ Mpc}^{-3}}\right)\nonumber\\\,{}\left(\frac{f_{\rm SF}}{0.8}\right)\,{}\left(\frac{600}{n_{\rm SC}}\right)\,{}\left( \frac{3500\text{ M}_\odot{}}{M_{\rm SC}}\right)\,{}\left(\frac{\sum_{\rm i} \tilde{t}_{\rm GW, i}^{-1}}{8\,{}{\rm Gyr^{-1}}}\right), 
\end{eqnarray}
where $\rho_{\rm SF}$ is the cosmological star formation rate density at redshift zero ($1.5 \times 10^{-2} \text{ M}_{\odot} \text{ yr}^{-1} \text{ Mpc}^{-3}$, \citealt{hopkins2006}), $t_{\rm life}$ is the average lifetime of a simulated SC  ($=100$ Myr), $f_{\rm SF}$ is the fraction of star formation that occurs in SCs (=0.8 from \citealt{lada2003}), $n_{\rm SC}$ is the number of simulated SCs, $M_{\rm SC}$ is the average mass of the SCs ($=3500 \,{}\text{M}_{\odot}$), 
and $\tilde{t}_{\rm GW, i}$ is the effective coalescence timescale of the i-th binary, as given by equation~\ref{eq:tildetgw}. Using the values of $\tilde{t}_{\rm GW}$ given in Table~\ref{tab:table2} for the three K16 merging systems, we find $R\sim{}5$ Gpc$^{-3}$ yr$^{-1}$, if we assume that all metallicities are equally likely. 

Applying the same equation (eq.~\ref{eq:drake}) to Z14 merging systems, we find  $R\sim{}11$ Gpc$^{-3}$ yr$^{-1}$ (\citealt{ziosi2014} give an estimate of $\sim{}3.5$ Gpc$^{-3}$ yr$^{-1}$ because they assume a model for the metallicity evolution in the local Universe and use a more approximate formula than equation~\ref{eq:drake}). Combining Z14 and K16 merging systems, the rate becomes $R\sim{}16$ Gpc$^{-3}$ yr$^{-1}$. Under the simplest assumptions, KL oscillations can increase the merger rate  by $\approx{}50$ per cent.

%\begin{table}
\begin{table}
\begin{center}
\caption{\label{tab:table2} List of the merging BH-BH systems and their associated parameters at the time of merging.}
 \leavevmode
 \begin{tabular}{cccccccc}
%\begin{tabular*}{\textwidth}{c @{\extracolsep{\fill}} ccccccc}
\hline
Ref. & $Z$ & $\tilde{t}_{\rm GW}$ & $m_1$ & $m_2$ & $m_{\rm c}$ & $m_{\rm t}$ & Bin. \\  
& (Z$_\odot$) & (Gyr) & (M$_{\odot}$)   & (M$_{\odot}$)  & (M$_{\odot}$) & (M$_\odot$) &Type  \\  
\hline
K16 & 0.01 & 0.16  & 61 & 20 & 30 & 5  & E\\ %31 IC_triple_111_001_4950+14950_4946_4004.txt  
K16 & 0.01 & 3.5   & 72 & 20 & 32 & 16 & E\\ %517 IC_triple_84_001_4676_4302_4946.txt
K16 & 0.1  & 0.5   & 39 & 18 & 23 & 15 & E\\ %270 IC_triple_183_010_4254_2115_989.txt
\hline
Z14 & 0.01 & 0.09 & 50 & 40 & 39 & -- & E \\%& Z001n154idsa1746b4480+14480\\
Z14 & 0.01 & 1.34 & 12 & 8  &  8 & -- & P \\%& Z001n052idsa3370b13370\\
Z14 & 0.01 & 1.76 & 9  & 9  & 8  & -- & P\\%& Z001n194idsa2870b12870\\
Z14 & 0.01 & 2.06 & 15 & 12 & 11 & -- & P \\%Z001n075idsa350b10350
Z14 & 0.1  & 0.20 & 10 & 5  & 6  & -- & E \\%& Z010n200idsa3555b180\\
Z14 & 0.1  & 0.67 & 11 & 7  & 8  & -- & P \\%& Z010n185idsa4340b14340\\
Z14 & 0.1  & 1.49 & 10 & 9  & 8  & -- & P\\%& Z010n162idsa2070b12070\\
\hline
A16a & $<0.5$ & $<14$ & $36.2^{+5.2}_{-3.8}$ & $29.1^{+3.7}_{-4.4}$ & $28.1^{+1.8}_{-1.5}$ & -- & U \\
A16b & -- & $<14$ & $14.2^{+8.3}_{-3.7}$ & $7.5^{+2.3}_{-2.3}$ & $8.9^{+0.3}_{-0.3}$ & -- & U \\
A16b & -- & $<14$ & $23^{+18}_{-6}$ & $13^{+4}_{-5}$ & $15.1^{+1.4}_{-1.1}$ & -- & U \\
\noalign{\vspace{0.1cm}}
\hline
\end{tabular}
%\end{tabular}
\begin{flushleft}
\footnotesize{Column 1 (Ref.): reference of the paper from which the system was taken. K16 means this paper; Z14 indicates \cite{ziosi2014};  A16a and A16b indicate \cite{abbott2016} and \cite{GW151226a}, respectively. The last three rows show the properties of the two GW events (GW150914, GW151226) and the third possible signal (LVT151012) detected by Advanced LIGO \citep{abbott2016,GW151226a}, for comparison. Column 2 ($Z$): metallicity; column 3 ($\tilde{t}_{\rm GW}$): effective coalescence time since the beginning of the  simulation (eq.~\ref{eq:tildetgw});  column 4 ($m_1$): mass of the primary member of the inner binary; column 5 ($m_2$): mass of the secondary member of the inner binary; column 6 ($m_{\rm c}$): chirp mass of the  inner binary; column 7 ($m_{\rm t}$): mass of the tertiary, when present; column 8 (Binary type): origin of the binary. `P' and `E' stand for primordial and exchanged binary, respectively. In the last column, `U' means unknown origin.}
\end{flushleft}
\end{center}
\end{table}
%\end{table}

We can also calculate the detection rate for Advanced LIGO/Virgo, $R_{\rm det}$, by extrapolating the merger rate to a volume set by the radius out to which the mergers produce a GW signal within the instrumental range of Advanced LIGO/Virgo (this range depends on the chirp mass of each binary):
\begin{multline}\label{eq:detec}
R_{\rm det} = \frac{4\pi}{3} \frac{t_{\rm life} \,\rho_{\rm SF}\, f_{\rm SF}}{n_{\rm SC}\,{} M_{\rm SC}} \,{} \left( \frac{d_{H}}{f} \right)^3\frac{1}{m_{\rm c, 10}^{15/6}} \sum_{\rm i} \frac{m_{\rm c, i}^{15/6}}{\tilde{t}_{\rm GW, i}},   
\end{multline}
where $f=2.26$ is a correction factor accounting for the random location on the sky and orientation of sources, as well as the non-uniform pattern of detector sensitivity \citep{finn1996,belczynski2013}, $m_{\rm c, 10}=8.7$ M$_\odot$ is the chirp mass of a BH-BH binary composed of two 10 M$_\odot$ BHs, $d_{H}=1$ Gpc is the approximate instrumental range of Advanced LIGO and Virgo \citep{abadie2010} for a binary with  $m_{\rm c}=m_{\rm c, 10}=8.7$ M$_\odot$, and $m_{\rm c, i}$ is the chirp mass of i-th merging BH-BH system.

Adopting the same values for $t_{\rm life}$, $\rho{}_{\rm SF}$, $f_{\rm SF}$, $n_{\rm SC}$ and $M_{\rm SC}$ as discussed before, and using $t_{\rm GW, i}$ and $m_{\rm c, i}$ from Table~\ref{tab:table2}, we find $R_{\rm det}\sim{}40$ yr$^{-1}$. For comparison, the detection rate of the seven Z14 merging systems, derived from eq.~\ref{eq:detec} is $R_{\rm det}\sim{}100$ yr$^{-1}$. Combining Z14 and K16 merging systems, the detection rate becomes $R_{\rm det}\sim{}140$ yr$^{-1}$. We note that  $d_{H}=1$ Gpc will be reached by Advanced LIGO approximately in the O2 run and that equation~\ref{eq:detec} assumes that our simulated mergers are homogeneously distributed in a sphere of $1$ Gpc. Thus, our estimate is likely an upper limit.

Given the low statistics, these results have large uncertainties and should be considered only as preliminary predictions of the importance of Kozai merging systems with respect to the other merging systems in young SCs. Thus, we find that including the effect of KL resonances can increase the detection rate by $\sim{}40$ per cent. This is a mild effect but is extremely important if we consider the properties of systems that merge by KL resonances. As we discussed in the previous Section, Kozai merging systems have generally higher masses than the other BH-BH systems. Thus, they can be detected at larger distances, and are consistent with the mass of the observed GW150914 event. Distinguishing between Kozai triggered mergers and mergers of primordial binaries can give us insights into the mechanisms that lead to the formation of a BH-BH binary.

\section{Summary}
HTs, i.e. triple systems where an inner binary is orbited by a tertiary body further out, are very common in the local Universe. If the orbital plane of the tertiary is inclined with respect to that of the inner binary, the system might undergo KL resonance: the eccentricity of the inner binary and the inclination between the two orbital planes start oscillating. Since the coalescence timescale of a double compact object binary strongly depends on its eccentricity, the periodic increase of eccentricity due to KL oscillations can affect the merger rate significantly.

In \cite{ziosi2014}, we simulated the formation and the dynamical evolution of double compact-object binaries in young SCs. We found that $\sim{}27$ per cent of all BH-BH and BH-NS binaries (corresponding to 570 systems) are  members of a HT  with another BH or, in few cases, with a NS. However, the algorithms adopted by \cite{ziosi2014} were not sufficiently accurate to study the development of KL oscillations in these 570 HTs. In this paper, we re-simulate the HTs found in \cite{ziosi2014} by using a new code based on the Mikkola's algorithmic regularization scheme. We also included the 2.5PN term. 

We find that $\sim{}88$ per cent of the simulated HTs develop KL oscillations. Less than $\sim{}2$ per cent of the simulated HTs undergo an exchange during the simulation, while the remaining $\sim{}10$ per cent of simulated HTs do not show significant KL oscillations or other interesting dynamical features.

In three runs (i.e. $\sim{}0.5$ per cent of all runs) KL oscillations lead to the merger of the inner binary, if the 2.5PN term is included. Even if this is a small sample to make any strong statements, we note that all three merging systems originated from dynamical exchanges (i.e. none comes from a primordial binary). In contrast, \cite{ziosi2014} reported seven systems merging in a Hubble time, 70 per cent of which were born from primordial binaries and only 30 per cent from dynamical exchanges. Moreover, most of the merging systems simulated by \cite{ziosi2014} have lower masses than the three systems merging by KL oscillations.  The BH masses of merging systems found by \cite{ziosi2014} are fairly consistent with those of  GW151226 (the second LIGO event), while the three systems merging  by KL oscillations have higher masses, similar to those of GW150914 (the first LIGO event).

Dynamical interactions trigger the formation of BH-BH binaries more massive than  BH-BH binaries originating from primordial binaries \citep{ziosi2014,morscher2015,giersz2015,mapelli2016}. However, most of these massive dynamically formed BH-BH binaries are not expected to merge, according to the results of \cite{ziosi2014}, who did not account for KL oscillations. In our paper, we show that %KL oscillations can boost the merger of such massive dynamically formed binaries.
%Thus, 
KL oscillations might enhance the merger rate of massive dynamically formed BH-BH binaries. This result is particularly important to interpret the formation channel of GW150914: while it is highly controversial whether GW150914 comes from a primordial binary or from a dynamically formed system (e.g. \citealt{LIGO2016b}), such high-mass BH-BH binary can result from the evolution of primordial binaries only with fine tuning (e.g. \citealt{belczynski2016,belczynski2016b,marchant2016}). Our simulations indicate that dynamical mechanisms (e.g. exchanges) trigger the formation of massive BH-BH binaries \citep{ziosi2014}, and KL oscillations can boost the merger rate of such massive dynamically formed BH-BH binaries.
 
%through a lot of fine tuning (e.g. \citealt{belczynski2016}). Our simulations suggest that KL oscillations can trigger the merger of particularly massive systems.

We estimate that the merger rate $R$ of BH-BH binaries from \cite{ziosi2014} is $R\sim{}11$ Gpc$^{-3}$ yr$^{-1}$, while the merger rate of the three systems merging by KL oscillations is $\sim{}5$ Gpc$^{-3}$ yr$^{-1}$. Combining these results, we find that a total merger rate  $R\sim{}16$ Gpc$^{-3}$ yr$^{-1}$ is expected from the simulations of \cite{ziosi2014} when accounting for Kozai resonances. While this result is affected by large uncertainties, it is fairly consistent with the constraints posed by GW150914  and GW151226 ($R\sim{}9-240$ Gpc$^{-3}$ yr$^{-1}$, \citealt{GW151226a}).

Our results suffer from a number of issues. First, we simulated the HTs in isolation. While we checked that none of the merging HTs is affected by strong perturbers during the $N$-body simulations of \cite{ziosi2014}, we cannot exclude that even  weak perturbations can prevent any of such systems from merging. A full $N$-body simulation with a regularization scheme is requested to check this issue. Moreover, we neglected non-dissipative PN terms. They do not contribute to GW decay but can induce precession of the inner binary and affect the dynamical evolution of the HT. Their contribution will be considered in a forthcoming study. Taking into account all these {\it caveats}, our results indicate that HT formation and Kozai resonance might have a crucial impact on the demographics of GW sources. In particular, KL oscillations can trigger the coalescence of  massive BH-BH systems in dynamically formed binaries, leading to an increase of the merger rate  by $\approx{}50$ per cent.

\section*{Acknowledgments}
We thank the anonymous referee for their useful comments, which improved this work significantly. We also thank Mauro D'Onofrio, Alessandro Bressan, Ugo Niccol\`o Di Carlo and Enrico Montanari for useful discussions. TK acknowledges financial support from the Erasmus+ programme. MM, MS and BMZ acknowledge financial support from the Italian Ministry of Education, University and Research (MIUR) through grant FIRB 2012 RBFR12PM1F, and from INAF through grant PRIN-2014-14. MM acknowledges financial support from the MERAC Foundation.
\bibliography{./bibliography}

\end{document}